\pdfoutput=1

\documentclass[11pt]{article}

\usepackage{acl}

\usepackage{times}
\usepackage{latexsym}

\usepackage[whole]{bxcjkjatype} 

\usepackage[T1]{fontenc}

\usepackage[utf8]{inputenc}

\usepackage{microtype}

\usepackage{inconsolata}

\usepackage{graphicx}

\usepackage{multirow}

\title{A Methodology for Identifying Evaluation Items for Practical Dialogue Systems Based on Business-Dialogue System Alignment Models}

\author{
  \textbf{Mikio Nakano\textsuperscript{1,3}},
  \textbf{Hironori Takeuchi\textsuperscript{2}},
  \textbf{Kazunori Komatani\textsuperscript{3}}
\\
\\
  \textsuperscript{1}C4A Research Institute, Inc., Setagaya, Tokyo, Japan \\
  \textsuperscript{2}Musashi University,  Nerima, Tokyo, Japan\\
  \textsuperscript{3}SANKEN, Osaka University, Ibaraki, Osaka, Japan\\
 \texttt{mikio.nakano@c4a.jp, h.takeuchi@cc.musashi.ac.jp}\\
 \texttt{komatani@sanken.osaka-u.ac.jp}
  }

\begin{document}
\maketitle
\begin{abstract}

This paper proposes a methodology for identifying evaluation items
for practical dialogue systems. Traditionally, user satisfaction and user experiences have been
the primary metrics for evaluating dialogue systems. However, there
are various other evaluation items to consider when developing and
operating practical dialogue systems, and such evaluation items are expected to lead to new research
topics. So far, there has been no methodology for identifying these
evaluation items. We propose identifying evaluation items based on 
business-dialogue system alignment models, which are applications of
business-IT alignment models used in the development and operation of practical IT systems. We also present a generic
model that facilitates the
construction of a business-dialogue system alignment model for each
dialogue system. 

\end{abstract}

\section{Introduction}

Traditionally, in the dialogue systems research community, user satisfaction \cite{walker-etal-1997-paradise,ultes:dd21,pan-etal-2022-user} and user experience \cite{clark:chi19,DBLP:journals/que/FolstadT21,Johnston2023,minato:ar23} have been widely used as metrics for evaluating dialogue systems. With recent advancements in dialogue system technology, particularly the development of large language models (LLMs), it has become possible to develop dialogue systems with high scores in these metrics \cite{hudecek-dusek-2023-large,iizuka2023clarifying}.

However, in developing and operating practical systems, it is necessary to consider various factors other than the aforementioned metrics. For instance, a chatbot using Retrieval-Augmented Generation (RAG) \cite{rag} can generate natural responses based on the contents of a database, but there is still a possibility of generating responses that are inconsistent with the database contents. Therefore, there are risks associated with using such a system for customer service. Additionally, when using an LLM on one's own hardware, substantial hardware resources are required, resulting in high running costs. Consequently, if the anticipated benefits do not exceed these costs, it is difficult to continue operating the system.

In addition to LLMs, various new technologies have been proposed for dialogue systems, but not all are used in practical systems. We suspect that one reason for this is the difference between the evaluation metrics used in the research community and those used to evaluate practical systems. So it is crucial to identify evaluation items for building and operating practical systems.

\newcite{Dybkjaer} and \newcite{McTear2004}  mention requirements for dialogue systems in explaining dialogue systems development life cycles. \newcite{McTear2004} discusses the need for considering requirements from not only users but also operators, but how to list all the requirements is not discussed.
\newcite{nakano:iwsds24} categorize evaluation items for dialogue systems from the system owner's perspective into benefits, costs, and risks, and they include items that do not have a positive correlation with user satisfaction or user experience. However, the methodology for identifying all evaluation items for individual dialogue systems has not been presented.

In this paper, we apply business-IT alignment models
\cite{hinkelmann2016new} to dialogue systems. Business-IT alignment
models are widely used to link business goals, business processes,
and applications to facilitate the examination and evaluation of
business systems by various stakeholders. We call the results of the
application of business-IT alignment models to dialogue systems {\bf
  Business-Dialogue System Alignment Models} (hereafter Business-DS Alignment Models). By applying these models
to individual dialogue systems to create a business-DS
alignment, it becomes possible to list evaluation items specific to
each dialogue system.

Furthermore, to facilitate the creation of the business-DS alignment model for an individual dialogue system, this paper proposes a {\em generic model for business-DS alignment}. By applying this generic model to individual dialogue systems, it is possible to create an alignment model tailored to each system, which can then be used to identify the corresponding evaluation items.

It should be noted that, while this paper uses the term {\em business}, it is not limited to the narrow sense of business. Instead, it encompasses all practical dialogue system development and operation. For example, the same analytical approach can be applied to systems developed and operated by non-profit organizations or local governments.

\section{Previous Work}

\subsection{Evaluating Dialogue Systems}

As previously mentioned, user satisfaction \cite{walker-etal-1997-paradise,pan-etal-2022-user,ultes:dd21} and user experience \cite{clark:chi19,DBLP:journals/que/FolstadT21,Johnston2023,minato:ar23} are commonly used metrics for evaluating dialogue systems. User satisfaction is measured by integrating factors such as the degree of task completion and the cost incurred by the user to achieve the task \cite{walker-etal-1997-paradise}. User experience is generally measured through subjective evaluations. Post-interaction surveys are often used to ask questions such as whether the interaction with the system was enjoyable or if the user would like to converse with the system again.

However, there are also studies addressing important factors that cannot be measured by these metrics alone. One such factor is development cost. Recent dialogue system technologies often utilize models trained with annotated data. Using active learning to achieve higher accuracy with a smaller amount of annotations is proposed \cite{asghar-etal-2017-deep, hiraoka2017active, tur2005combining}. Additionally, end-to-end learning for building dialogue systems \cite{lowe2017training, wen-etal-2017-network} can reduce development costs by eliminating the need for annotations.
Furthermore, research is also being conducted to reduce hardware costs during operation \cite{DBLP:journals/nca/PandeleaRYGC22}.

In addition, recent neural dialogue generation and dialogue systems using large language models may include offensive or discriminatory language in their utterances. Methods for avoiding such utterances are also proposed \cite{xu2021recipes, sun-etal-2022-safety, ziems-etal-2022-moral, Henderson:AIES18}.

However, no methodology has been proposed to identify all the items to evaluate when developing and operating practical dialogue systems.

\subsection{Business-IT Alignment Model}

To identify all the evaluation items, it is necessary for various stakeholders involved in the development and operation to overview and evaluate the project from their respective perspectives. This requires a comprehensive view of the entire project.

In the context of IT systems in general, not limited to dialogue systems, discussing systems from both managerial and developmental viewpoints is referred to as {\em business-IT alignment}. To achieve this, the relationships between business goals, business processes, and applications are represented in what is called a {\em business-IT alignment model}.

In a business-IT alignment model, it is possible to represent not only the IT system itself but also its development and operation. \newcite{DBLP:conf/icexss/VicenteGS13} created models for operation and \newcite {DBLP:journals/sosym/MayerAGFGW19} created models for risks.

There is also research on modeling business-IT alignment for AI service systems that use machine learning \cite{DBLP:conf/kesidt/TakeuchiY19}. Additionally, meta-models that integrate multiple models related to AI service systems have been proposed \cite{HJati2024, HTakeuchiFGCS2024}.

However, dialogue systems are different from typical AI service systems in that they intensively interact with humans. Therefore, the aforementioned models cannot be directly applied to dialogue systems.

\section{Proposed Methodology}

\subsection{Overview}

We propose a methodology in which various stakeholders involved in a
dialogue system development and operation project can overview and
evaluate the project from their respective perspectives by
constructing a business-DS alignment model. Based on this
model, we identify comprehensively the evaluation items.

A business-dialogue system alignment model consists of the services provided by the dialogue system, values, risks, and costs. Each of these components is broken down into finer elements and represented using a modeling language called ArchiMate \cite{josey2019archimate}. By further integrating these elements and expressing the relationships between them, the overall model can be represented. This allows for the enumeration of the values, risks, and costs associated with the target dialogue system.

\begin{figure}[t]
  \centering
  \includegraphics[width=.95\linewidth]{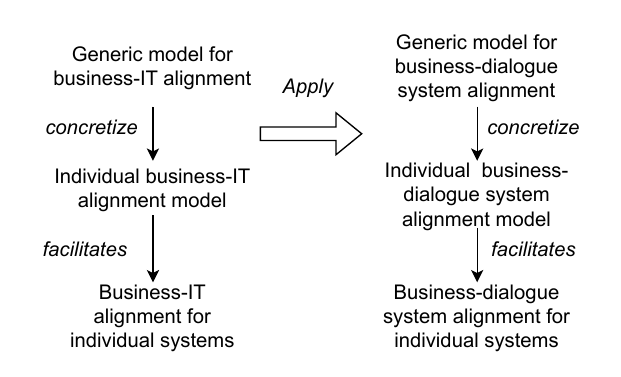}  
\caption{Relationships among the business-IT and business-DS alignment models.}
\label{models}
\end{figure}

However, constructing a business-DS alignment model from scratch is difficult for researchers in the dialogue system community. Therefore, we propose a generic model for business-DS alignment. Applying this generic model to individual dialogue systems makes it easy to create an alignment model tailored to each system, which can then be used to list evaluation items.
Figure~\ref{models} illustrates the relationship among business-IT
alignment models and business-DS alignment models.

\begin{table}[t]

  \centering
  \includegraphics[width=.95\linewidth]{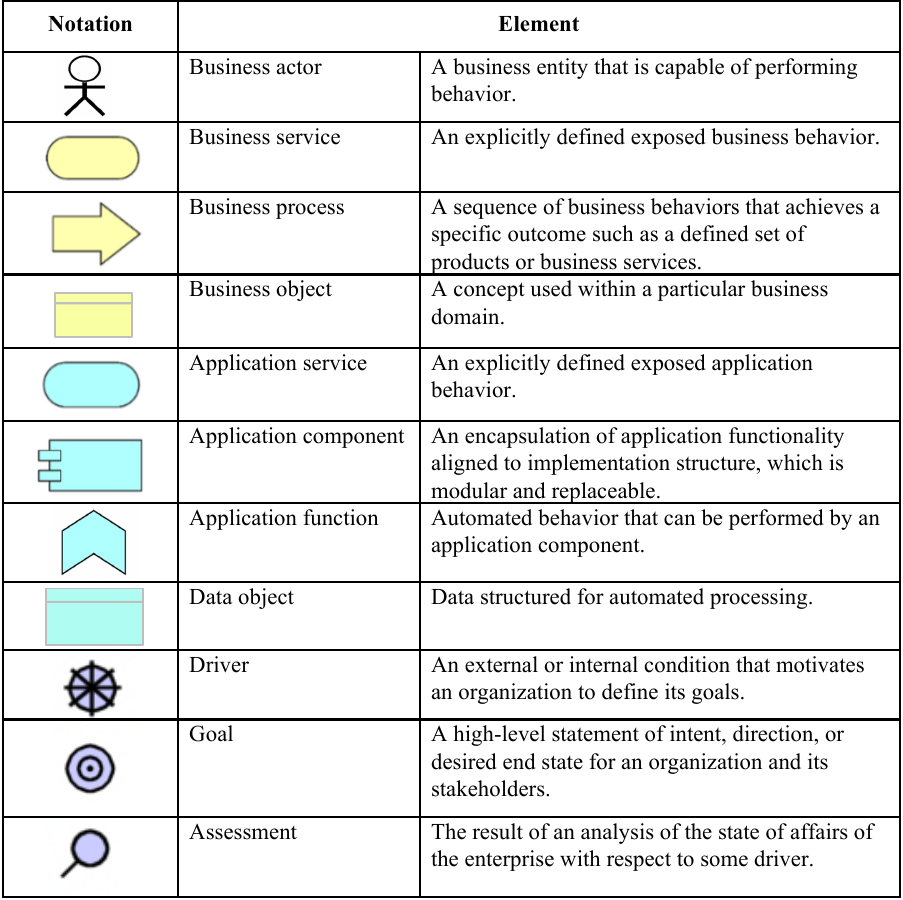}
  \caption{ArchiMate elements.}
  \label{elements}
\end{table}

\begin{table}[t]

  \centering
  \includegraphics[width=.95\linewidth]{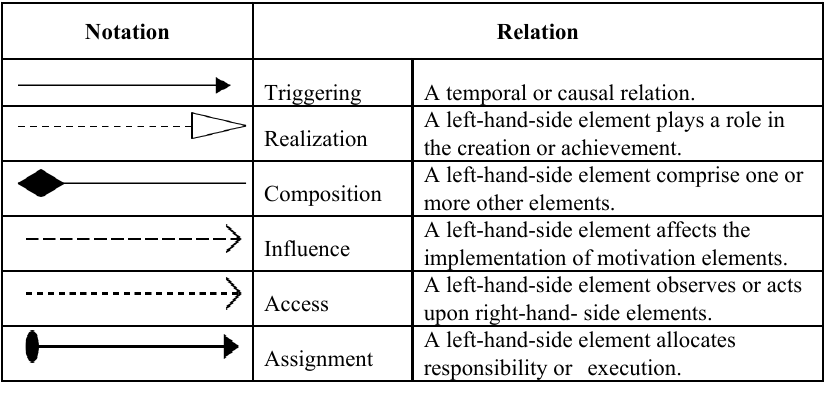}
    \caption{ArchiMate relations.}
  \label{relations}
\end{table}

\subsection{Generic Model for Business-Dialogue System Alignment}
\label{sec:genericmodel}

The generic model for business-DS alignment consists of the generic model of values, the generic model of risks, the generic model of costs, and the generic model of the services provided by dialogue systems (hereafter, we simply call this {\em the generic model of dialogue systems}).
We illustrate these using ArchiMate. The explanations of the ArchiMate elements and relationships are shown in Tables \ref{elements} and \ref{relations}, respectively.

\begin{figure*}[t]
  \centering
  \includegraphics[width=.75\linewidth]{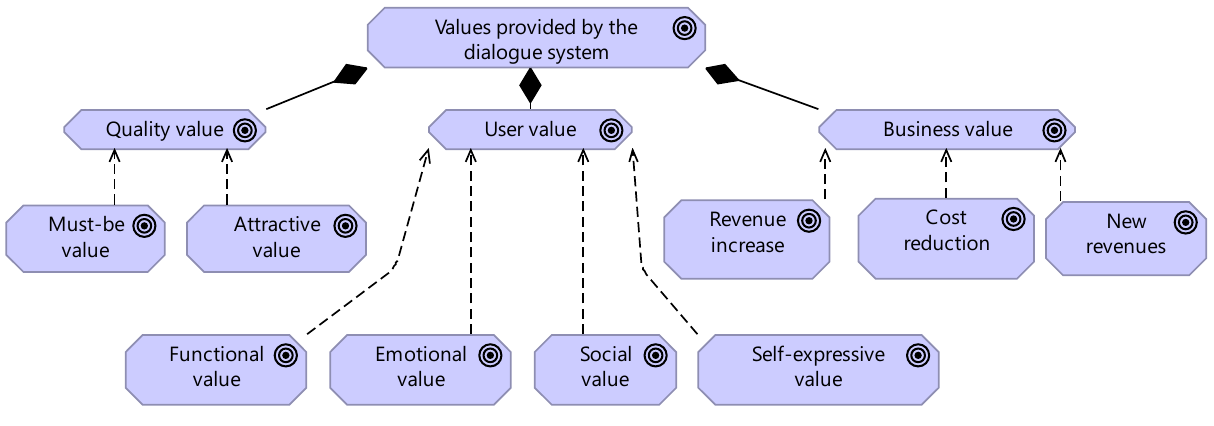}
  \caption{Generic model of values.}
  \label{value}
\end{figure*}

\subsubsection{Generic Model of Values}
\label{sec:value}

The values of dialogue systems are defined from various perspectives. We
consider that it consists of {\em user value}, {\em quality value},
and {\em business value}, and further enumerate their sub-elements.
Figure~\ref{value} is the ArchiMate illustration for these.

\paragraph{User value}

User value refers to the benefits that users obtain. Following \newcite{aaker2014aaker}, we consider the following four elements as components of the value model:

\begin{itemize}

\item {\bf Functional value}: The utility obtained from the functions of the service: e.g., achieving tasks or effectively practicing dialogue.

\item {\bf Emotional value}: The special emotions brought about by the process and experience of using the service: e.g., enjoying the conversation.

\item {\bf Self-expressive value}: The state where users can express their ideal selves through the use of the service: e.g., feeling satisfied with one's ability to effectively use the dialogue system.

\item {\bf Social value}: The identity or sense of belonging gained from using the service: e.g., feeling satisfied being part of a group that uses the same dialogue system.

\end{itemize}

\paragraph{Quality value}

Quality value refers to the value that users obtain from the high quality of the service. Based on the quality model called the Kano Model \cite{kano1984attractive,mikulic2011critical}, we decompose quality value into the following elements.

\begin{itemize}

\item {\bf Must-be value}: This value leads to dissatisfaction if not fulfilled but does not significantly increase satisfaction when fulfilled. 
In the context of dialogue systems, this includes the ability to complete tasks reliably and the system not crashing.

\item {\bf Attractive value}: This value does not cause dissatisfaction if not fulfilled, but significantly increases satisfaction when it is. For dialogue systems, this includes the ability to engage in natural, human-like conversation, such as fluency and appropriate timing and prosody.

\end{itemize}

\paragraph{Business value}

Business value refers to the value obtained by the operators or owners of the dialogue system. The following three elements are considered sub-components:

\begin{itemize}

\item {\bf Revenue increase}: This includes the increase in sales of products incorporating the dialogue system and the increase in sales of products recommended by the dialogue system.

\item {\bf Cost reduction}: This refers to the reduction in labor costs achieved by replacing tasks previously performed by humans with the dialogue system.

\item {\bf New revenue}: This includes revenue from service fees for using the dialogue system, income from displaying advertisements to dialogue systems users, and revenue from selling collected dialogue data.

\end{itemize}

Here, quality value demonstrates attributes such as ``whether not providing it poses a risk'' or ``whether providing it leads to opportunities.'' On the other hand, business value can be seen as what the provider gains in exchange for delivering user value \cite{perri}.

Here, we have listed quality value, user value, and business value in parallel. However, enhancing quality value and user value can lead to an increase in the number of users and usage frequency, which in turn may lead to revenue increase, cost reduction, and new revenues. These relationships vary depending on the individual system.

\begin{figure*}[t]
  \centering
  \includegraphics[width=.65\linewidth]{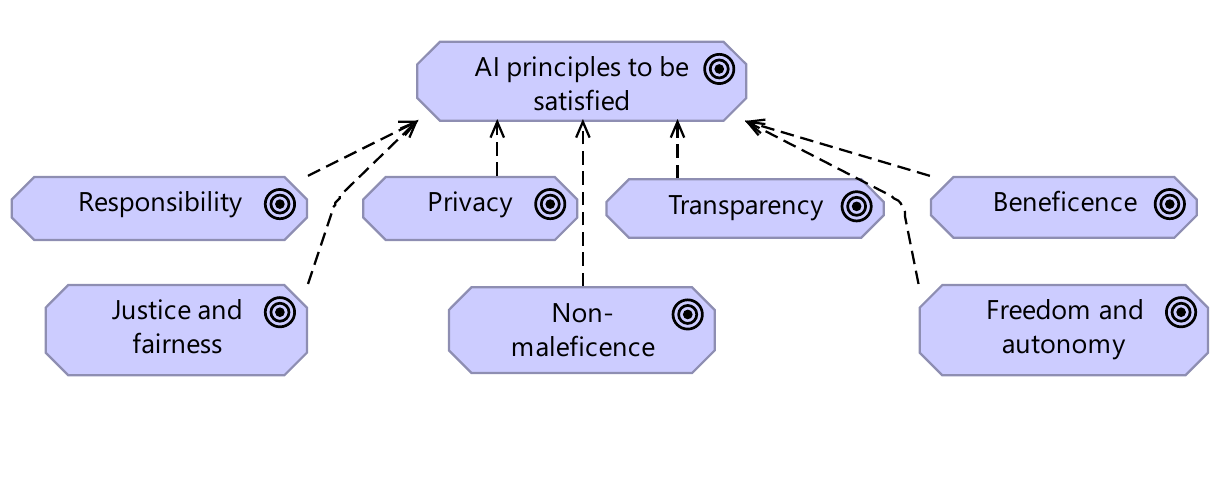}
  \caption{Generic model of risks. Not satisfying the AI principles causes risks.}
  \label{risk}
\end{figure*}

Note that we do not limit the dialogue systems targeted in this study to task-oriented dialogue systems. Non-task-oriented dialogue systems can also have various values. For example, in the case of a system that allows users to chat with a well-known character \cite{akama-etal-2017-generating, han-etal-2022-meet}, users can gain emotional value by enjoying casual conversations. Additionally, since the system can promote the character, the system owner can achieve a revenue increase.

\subsubsection{Generic Model of Risks}
\label{sec:risks}

In recent years, there have been many concerns about the risks associated with AI, including generative AI. In this context, principles for the societal implementation of AI are being considered not only by academic organizations but also by national and international institutions. This study views the failure to adhere to these principles as a risk.

Many principles have been established as guidelines, but they vary in granularity and comprehensiveness, and comparisons are being made \cite{jobin2019global}. In our study, the principles mentioned in more than one-third of the 84 guidelines investigated by \newcite{jobin2019global} are considered components of risk, and we apply these principles to dialogue systems.

\begin{description}

\item [Transparency:] The dialogue system can explain why it behaved in a certain way.

\item [Justice and fairness:] It does not make utterances based on biased thinking.

\item [Non-maleficence:] There is no risk of generating defamatory utterances, producing incorrect utterances, or copyright violation.

\item [Responsibility:] Responsibility is clearly assigned when problems arise.

\item [Privacy:] There is no risk of leakage of personal information, speech, or facial images contained in the dialogue content.

\item [Beneficence:] The dialogue system has a positive impact on users and society.

\item [Freedom and autonomy:] There is no risk of being used for criminal purposes.

\end{description}

When developing or operating dialogue systems, if there is a possibility that these principles could be compromised, it is considered to be a risk.

Figure~\ref{risk} illustrates this generic model of risks.

\begin{figure}[t]
  \centering
  \includegraphics[width=.95\linewidth]{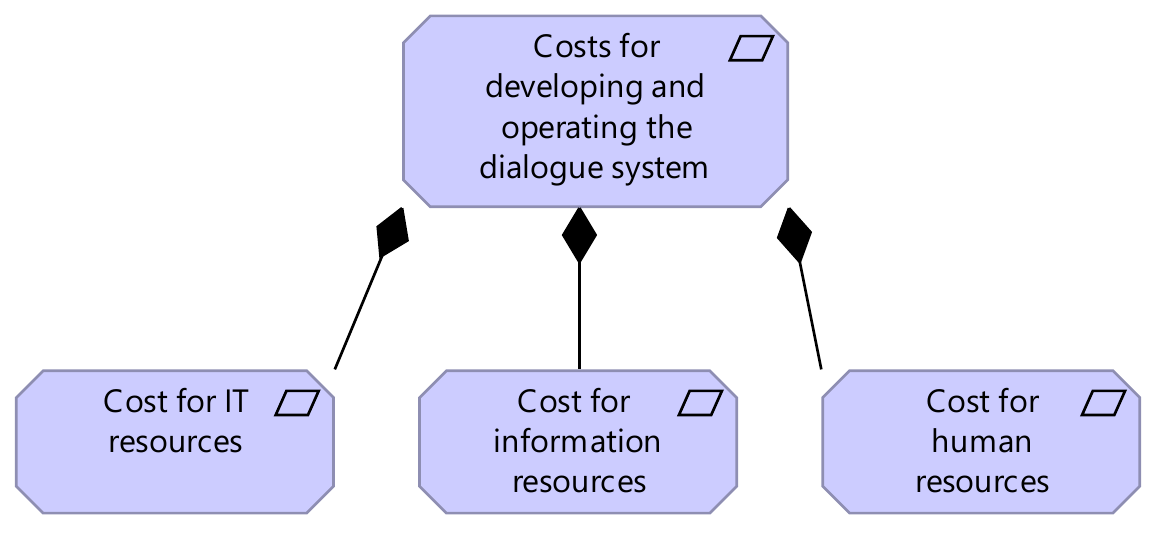}
  \caption{Generic model of costs.}
  \label{cost}
\end{figure}

\subsubsection{Generic Model of Costs}
\label{sec:costs}

In the practical implementation of any system, not limited to dialogue systems, development and operational costs are required. These costs can be broken down as follows:

\begin{description}

\item [Cost for human resources:] This includes human resources for initial system development, system testing, system modifications after the start of operation, and human resources for handling issues and troubleshooting.

\item [Cost for information resources:] This involves the creation of annotated data for model building, and the creation of data used as references for writing rules.

\item [Cost for IT resources:] This includes computing resources needed for initial system development, server usage fees, external API service usage fees, and application registration fees.

\end{description}

Figure~\ref{cost} illustrates this generic model of costs.

\begin{figure*}[t]
  \centering
  \includegraphics[width=.99\linewidth]{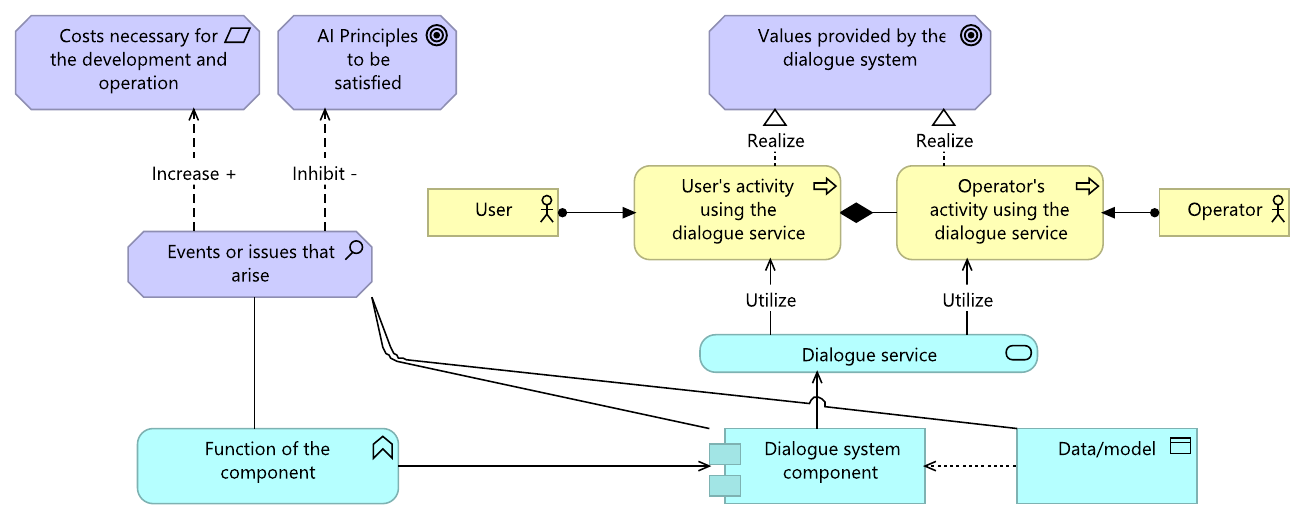}
  \caption{Generic model of dialogue systems. A solid line without direction denotes a general relationship.}
  \label{ds}
\end{figure*}

\subsubsection{Generic Model of Dialogue Systems}

Below we enumerate the elements related to a dialogue system. This is based on the AI service system description by \newcite{HTakeuchiFGCS2024}.

\begin{description}

\item[User:] The user of the dialogue system.

\item[Operator:] The person or entity operating or owning the dialogue system.

\item[User activities using the dialogue system:] Activities performed by the user using the dialogue system, such as performing tasks, practicing having a conversation, and enjoying a conversation.

\item[Operator activities using the dialogue system:] Activities performed by the operator using the dialogue system, such as providing information and obtaining information from users.

\item[Dialogue services:] Services provided by the dialogue system, such as providing information at any time and providing the joy of conversation.

\item[Dialogue system components:] Components within the dialogue system, such as language understanding component, dialogue management component, and information search component.

\item[Component functions:] Functions of the dialogue system components, such as language understanding, dialogue management, and information search.

\item[Data/models:] Models used by dialogue system components and the data to train these models, such as language understanding model and training data for it.

\item[Observed events/issues:] Possible events or issues regarding data/models, application components, or functions, such that annotated data for language model training is necessary and that the language generation component might generate incorrect statements.

\end{description}

Figure~\ref{ds} illustrates this generic model of dialogue systems.

\subsection{Creating a Business-Dialogue System Alignment Model and Identifying Evaluation Items}
\label{sec:eval}

To create a business-DS alignment model, we will apply the
general model described in Section~\ref{sec:genericmodel} to the target dialogue system.
In practice, each dialogue system will be represented using ArchiMate,
illustrating its relationships with value, cost, and risk
elements. Elements not related to these will be excluded. 

\begin{figure}[t]
  \centering
  \includegraphics[width=.95\linewidth]{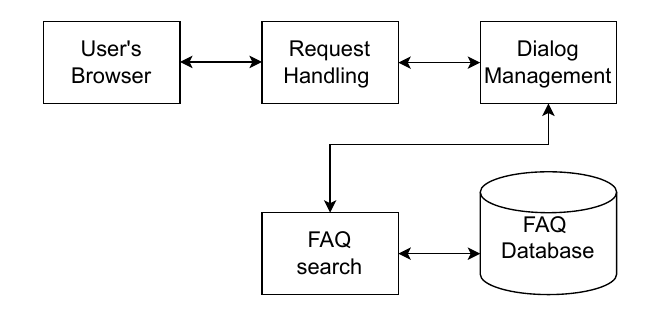}
\caption{Architecture of the FAQ chatbot as a case study.}
\label{faq-chatbot-arch}
\end{figure}

In the explanation below, we use a simple FAQ (Frequently-Asked Questions) chatbot as a case study. This chatbot uses an FAQ database containing question-and-answer pairs to respond to user queries via text input and output. It performs example-based question answering \cite{banchs-li-2012-iris,inaba-takahashi-2016-neural}. The system operates on a server, and users access it through a browser without entering a user ID. The chatbot comprises a web server for handling requests, a simple dialogue management module based on a state transition model, and an FAQ search module, as shown in Figure~\ref{faq-chatbot-arch}. The dialogue management module generates initial responses and handles situations where no FAQ match is found. The FAQ search module uses Sentence-BERT \cite{reimers2019sentencebert} to match the input sentence with example questions, extracts the relevant FAQ, and returns it to the dialogue management module.

\begin{figure*}[t]
  \centering
  \includegraphics[width=.95\linewidth]{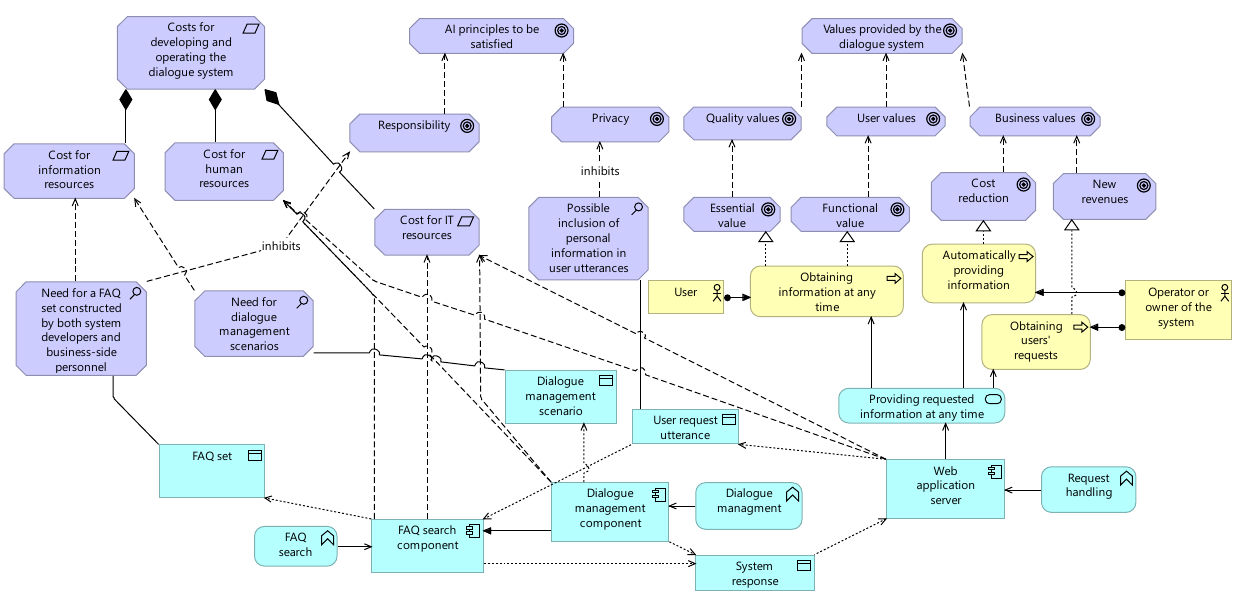}
  \caption{Business-dialogue system alignment model for FAQ Chatbot.}
  \label{faq-chatbot-alignment-model}
\end{figure*}

We first tailor the generic model of dialogue systems to the target system (Figure~\ref{faq-chatbot-alignment-model}).
In the case of the FAQ chatbot, it becomes as follows:

\begin{itemize}

\item {\em User} is the user of the dialogue system to seek information.

\item {\em Operator} is the operator or the owner of the system who provides information.

\item {\em User activity using the dialogue system} is obtaining information at any time.

\item {\em Operator activities using the dialogue system} are automatically providing information and obtaining users' requests.

\item {\em Dialogue service} is a service that provides information at any time.

\item {\em Dialogue system components} are web application server, dialogue management component, and 
  FAQ search component.

\item {\em Component functions} are request handling, dialogue management, and FAQ search.

\item {\em Data/models} are user request utterances, system responses, the dialogue management scenario, and the FAQ set.

\item {\em Observed events/issues} are the need for a FAQ set, the need for dialogue management scenarios, and the possible inclusion of personal information in user utterances.

\end{itemize}

Then these are linked to the elements of values, risks, and costs by the following steps.

\begin{itemize}
\item [(1)] Derive costs from observed events/issues in the development, operation, and usage of each component. 

 In the case of the FAQ chatbot, costs are required for developing and operating each component. Additionally, the need for a FAQ set and dialogue management scenarios incurs development and maintenance costs. 

\item [(2)] Identify principles that are hindered by observed events in the development, operation, and usage of each component as risks.

  In the case of the FAQ chatbot, the inclusion of personal information in user utterances poses a risk to privacy protection. On the contrary, since responses are pre-written in the FAQ database, the risk of incorrect answers, biased responses, or responses containing slander is low. Also, since the creation of the FAQ set involves cooperation between dialogue system developers/operators and business-side personnel, there is a risk of unclear responsibility for the content.

\item [(3)] Identify business value from activities associated with the dialogue system development operators.

  In the case of the FAQ chatbot, automating information providing reduces labor costs. Additionally, analyzing user requests can reveal user needs, leading to new revenue opportunities.

\item [(4)] Identify user value from user activities using the dialogue system and the business value influenced by that user value.

  In the case of the FAQ chatbot, the ability to obtain information provides functional value to the user.

\item [(5)] Identify quality value from user activities using the dialogue system and the business value influenced by that user value.

  In the case of the FAQ chatbot, the ability to obtain information at any time without service interruption provides essential value to the user.

\end{itemize}

In this way, the values, risks, and costs of individual dialogue systems are enumerated and identified as evaluation items. The resulting business-DS alignment model for the FAQ chatbot written in ArchiMate is shown in Figure~\ref{faq-chatbot-alignment-model}.

Additional case studies can be found in Appendix~\ref{addtional_case_studies}.

\section{Limitations and Discussion}

Although the case studies suggested that our approach is promising, there may be values, risks, and costs that have not been considered, necessitating continuous review. Particularly with advancements in technology like LLMs, which enable more natural conversations, new risks that were previously unconsidered may arise.

As stated earlier, academic research has often used user satisfaction and user experience as evaluation metrics. Roughly speaking, user satisfaction relates to functional value, self-expressive value, and social value.
User experience relates to emotional value, must-be value, attractive value, non-maleficence, justice and fairness, and transparency. Our analysis identified evaluation items beyond these, so it became possible to consider user satisfaction, user experience, and other evaluation items all at once. We hope this leads to new research themes.

In planning the actual system development, it is necessary to balance values, risks, and costs. For example, while showing many advertisements might increase business value, it could decrease emotional value and pose risks to hinder non-maleficence. Similarly, using a low-performance model to reduce costs can decrease must-be value. A balanced system design considering all evaluation items is necessary, and our approach enables such a balanced design by identifying evaluation items from various perspectives.

In some cases, it is desirable to integrate these evaluation items into a single-dimensional evaluation scale. However, the prioritization of these items must be determined by the consensus of various stakeholders, including the system owner. We hope business-DS alignment models help the facilitation among the stakeholders.

The evaluation items obtained using the methodology proposed in this paper do not necessarily allow for a quantitative assessment of dialogue systems. However, in many cases, various IT-related technologies are proposed and utilized without quantitative evaluation. 
In addition, focusing only on quantifiable evaluation items and ignoring other items have the risk of falling into the well-known {\em McNamara fallacy} (also known as the {\em quantitative fallacy}). 
We believe that instead of focusing solely on fields where quantitative evaluation is feasible through small-scale experiments, dialogue system researchers should also consider evaluation items that are difficult to quantify. This approach may lead to the development of more practical technologies.

While it is practically impossible to quantitatively demonstrate the superiority of our methodology, we aim to showcase its effectiveness by applying it to the development of a variety of practical dialogue systems and evaluating it from multiple perspectives.

Business-IT alignment models on which our methodology is based may not be familiar to dialogue system engineers, making it potentially challenging to construct a business-DS alignment model. Therefore, we believe it is effective to present a simpler model. As an alternative approach, it is also possible to consider developing human resources who can construct business-DS alignment models while communicating with various stakeholders.

\section{Concluding Remarks}

This paper proposed a methodology to identify evaluation items for dialogue systems based on business-DS alignment models. Although the methodology presented in this paper needs improvement through more case studies. 
Nevertheless, we believe that it serves as a useful first step.

Besides the future work already mentioned, We plan to analyze the issues that prevent commercializing systems in the research stage.

\section*{Acknowledgments}

This work was partly supported by JSPS KAKENHI Grant Number JP22H00536.


\bibliography{iwsds25_eval}

\clearpage

\appendix

\section{Additional Case Studies}
\label{addtional_case_studies}

\subsection{Dialogue Systems Analyzed}

In addition to the FAQ chatbot that was analyzed in Section~\ref{sec:eval}, we analyzed the system listed below.
We selected these systems because they
are already in commercial service or close to practical use.
Note that we do not assume the same settings as the systems referenced in the literature.

\paragraph{Speech-based assistant on smartphones}

This works as an embedded application of smartphones and performs question answering, controlling applications, and other tasks like Apple's iPhone Siri \cite{bellegarda2013spoken}.
The input modality is speech and the output modalities are speech, displaying on the smartphone, and application control.
It uses proprietary speech recognition.
Wake words are recognized on the device and other user utterances are recognized on the server.
It also uses proprietary server-based language understanding using BERT or others.
Dialogue management and response generation are rule-based and run on the server.
Speech synthesis is device-embedded.

\paragraph{Job interview practice system}

This system is designed for practicing job interviews \cite{Inoue2021,Yu2019,su18_interspeech} by interacting with a virtual agent.
The system operates on a server and is accessed via a browser.
The input modalities are speech and facial images, and the output modalities are speech and virtual agents.
It uses commercial server-based speech recognition and speech synthesis.
Language understanding, dialogue management, and language generation use an API-based commercial LLM service (such as OpenAI's ChatGPT\footnote{\url{https://openai.com/index/chatgpt/}}).
The virtual agent runs on the browser.

\paragraph{Interview dialogue system for understanding user status}

This is a virtual agent dialogue system designed to engage with users, asking about their lifestyle and health status while conversing with them \cite{10.5555/2615731.2617415,asao-etal-2020-understanding}. To ensure continuous use, the system aims to make the dialogues enjoyable for the users \cite{kobori-etal-2016-small}. The system operates on a server and is accessed through a browser. Input modalities are speech and facial images and the output modalities are speech and virtual agents. It uses server-based commercial speech recognition and language understanding, and device-embedded speech synthesis. It also uses scenario-based dialogue management running on the server. The virtual agent runs on a browser.

\paragraph{Conversational recommender system}

This system engages in dialogue to elicit user preferences and experience \cite{zeng-etal-2023-question}, and based on this information, recommends products \cite{10.1145/3453154,GAO2021100}. It operates on a server. The input and output modality is text.
It uses a crowd service for language understanding and state transition model-based dialogue management (e.g., Google Dialogflow\footnote{\url{https://cloud.google.com/dialogflow}}).

\subsection{Evaluation Items for Example Dialogue Systems}

Table~\ref{case-study} shows the elements of the generic model for business-DS alignment and their relation to each example system. 
The factors listed under ``common to all system'' are those shared by all systems.
We show this table instead of the comprehensive ArchiMate representations for simplicity.

Relatively minor risks have been omitted. For instance, even if rule-based utterance generation is used, there is a possibility that the person writing the rules might create biased or offensive utterance templates. However, this risk is generally low because checks are usually conducted before the system is deployed.

In contrast, response generation using LLMs carries a higher risk because it cannot be pre-checked. However, compared to other applications, job interview practice systems have relatively low actual harm even if the LLM generates inappropriate utterances. Considering the development cost, using an LLM is reasonable.

These case studies have suggested that, based on the business-DS alignment models, it is possible to identify the costs, risks, and values of individual dialogue systems. They also allow for highlighting potential issues and comparing systems from various perspectives.

\begin{table*}[tbp]

  \centering
  \includegraphics[width=.99\linewidth]{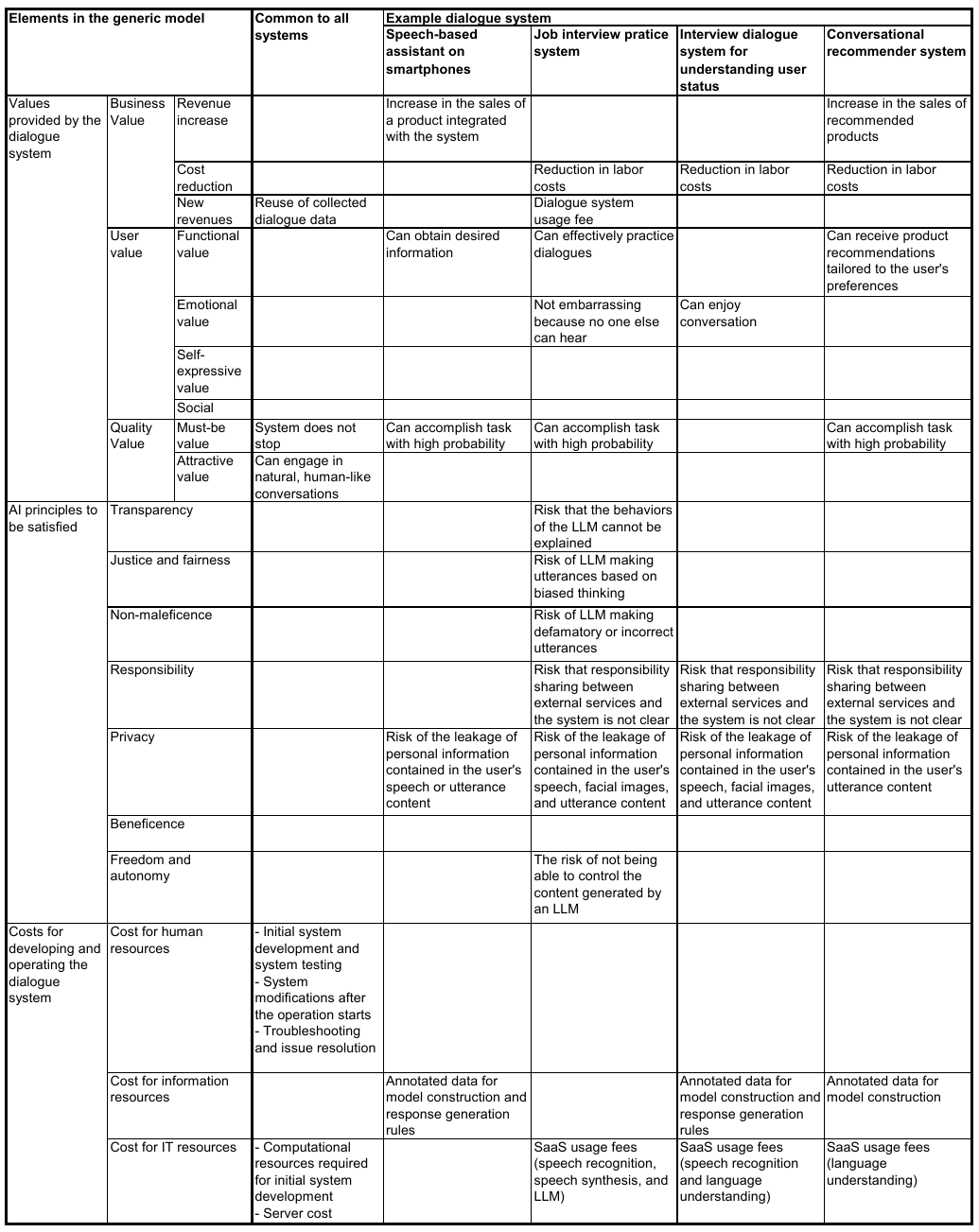}
    \caption{Evaluation items for example dialogue systems.}
  \label{case-study}
\end{table*}

\end{document}